\documentclass[onecollarge,natbib]{svjour2}
\bibpunct{[}{]}{;}{n}{}{,} 
\smartqed  
\usepackage{graphicx}
\journalname{Few-Body Systems (EFB22)}
\begin{document}

\title{Few-body Physics in a Many-body World
\thanks{The work of the author reported here was supported
by the Sapere Aude program under the Danish
Council for Independent Research.}}

\author{Nikolaj Thomas Zinner}

\institute{
              Department of Physics and Astronomy \\
              Aarhus University \\
              Ny Munkegade 120 \\
              DK-8000 Aarhus C, Denmark
              Tel.: +45-87155619\\
              \email{zinner@phys.au.dk}           
}

\date{Received: date / Accepted: date}

\maketitle

\begin{abstract}
The study of quantum mechanical few-body systems is a century old pursuit relevant to 
countless subfields of physics. While the two-body problem is generally considered to 
be well-understood theoretically and numerically, venturing to three or more bodies 
brings about complications but also a host of interesting phenomena. In recent years, 
the cooling and trapping of atoms and molecules has shown great promise to provide a 
highly controllable environment to study few-body physics. However, as is true for many
systems where few-body effects play an important role the few-body states are not 
isolated from their many-body environment. An interesting question then becomes 
if or (more precisely) when we should consider few-body states as effectively isolated 
and when we have to take the coupling to the environment into account. 
Using some simple, yet non-trivial, examples I will try to suggest possible 
approaches to this line of research.
\end{abstract}

\section{Introduction}
\label{intro}
Bound states of atomic particles are synonymous with the birth of modern quantum mechanics
through the pursuit of a solution to the Hydrogen atom and later more complicated 
atoms and molecules. From a mathematical point of view, bound states are studied 
by establishing criteria a given potential must satisfy in order for the system
to allow a bound state. Here we refer to a bound state in the usual sense, i.e. 
a state of negative energy with a normalizable wave function. Already in introductory
quantum mechanics classes we learn that the answer to this question depends very
strongly on dimensionality. In one dimension (1D), any amount of attraction will 
bind a particle, or more precisely, if the integral of the potential in all 
space (a line in this case) is non-positive. In two dimensions (2D) the situation 
is similar for the case of weak potentials since a bound 
state occurs when the integral of the potential in the plane is non-positive \cite{simon1976}.
However, the formula for the binding energy is very different from the case of 1D \cite{artem2011}.
In 3D this is no longer 
true. The simple example of an attractive square well potential demonstrates
this fact as a finite depth is required to obtain a bound state. Intuitively
this suggests that binding of two particles is harder to achieve in 3D. 
With this in mind, it must have been a bit of shock when Vitaly Efimov 
in 1970 announced that he had found an infinite number of three-body 
bound states in a 3D system with three bosonic particles in the limit where
each two-body subsystem has a bound state with zero binding energy 
\cite{efimov1970}. A flurry of theoretical 
interest followed and it was shown that it happens only in 3D and 
not in pure 1D or 2D \cite{tjon1979,nielsen1997,nielsen2001} 
(mixed dimensional examples have been found, see \cite{nishida2011}).
Efimov's finding is a key insight into the field of few-body physics as it
demonstrates the intricacies beyond two-particle systems.

Alas, in spite of many efforts to verify the predictions of Efimov in 
nuclear physics experiments no unambiguous signals have been found \cite{jensen2004}.
The observation was instead made in cold alkali atomic gases \cite{kraemer2006} 
about a decade after the production of the first atomic condensates \cite{davis1995,anderson1995,bradley1995}.
It was made possible by the advanced controllability of interactions that 
these systems provide through the use of Feshbach resonances \cite{inouye1998,chin2010}.
These resonances can be used to give the system an effective short-range interaction 
with a tunable strengh, or more precisly a tunable scattering length, $a$. 
In this notation, the Efimov effect occurs in the limit where $|a|\to\infty$, 
i.e. exactly on the Feshbach resonance. Since no quantity can depend on $a$
when it diverges, this is called the universal regime. An illustrative 
example of what this means can be seen in a (homogeneous) Fermi gas; when $|a|\to\infty$
the only scale left is the density (or Fermi wave vector), ergo the total 
energy of the system must be proportional to some constant times the 
Fermi energy (see \cite{giorgini2008} for a review of the Fermi gas
in the strongly interaction regime).

\begin{figure*}
\centering
\includegraphics[scale=0.4]{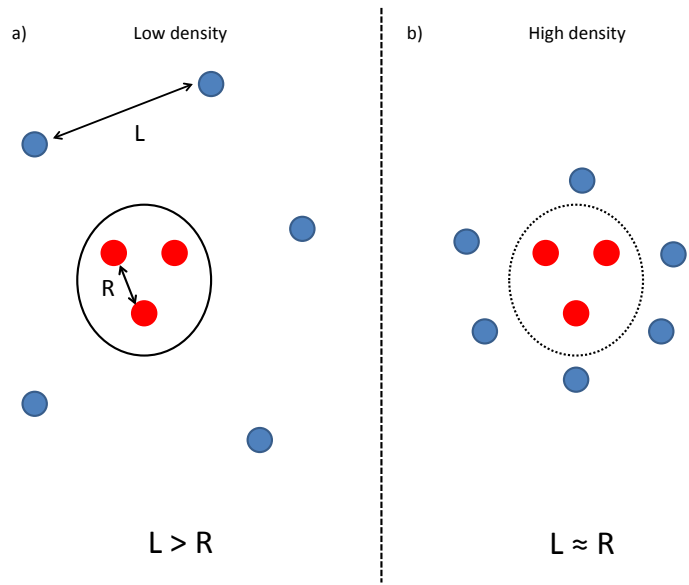}
\caption{(Left panel) An Efimov three-body state (three (red) balls surrounded by solid (black) circle) in an 
environment where the density is low ($L>R$) so that all other particles ((blue) balls outside the circle) are far away on average. (Right panel)
At higher densities ($L\sim R$), the interparticle distance in the total system becomes comparable to the interparticle in the 
three-body bound state and we expect a modification of its binding energy and/or character.}
\label{fig1}     
\end{figure*}

A question now arises at the 
borderline of few- and many-body physics. The Efimov effect is 
observed in many cold atomic gas experiments as a resonant peak 
in loss rates. This comes about since the presence of bound states
modifies the recombination rates as function of temperature and 
for given value of the probability to populate tightly-bound 
molecular two-body bound states 
which are abundantly present in alkali systems and cause losses
from the trap \cite{braaten2006,sorensen2013}. But what about the density of 
particles? Naturally, the loss rate will scale with the density
since shorter interparticle distances increase the probability
for reactions. The density could, however, also have an effect
of the position of the resonance peaks since the Efimov states
may feel the presence of other particles in their surroundings 
and endure a shift in binding energy. At low density we do not
expect this to happen (left panel in Fig.~\ref{fig1}), 
but as the density increases there should 
be a point at which the interparticle distance in the many-body
system becomes comparable to the interparticle distance in the 
few-body state (right panel in Fig.~\ref{fig1}) and the 
characteristics of the few-body state should change; it is 
no longer an isolated state. Below we discuss examples
of how to approach this issue when the background is a 
degenerate Fermi or Bose gas.

\section{Efimov states in a Fermi gas}
\label{sec:1}
\begin{figure*}
\centering
\includegraphics[scale=0.4]{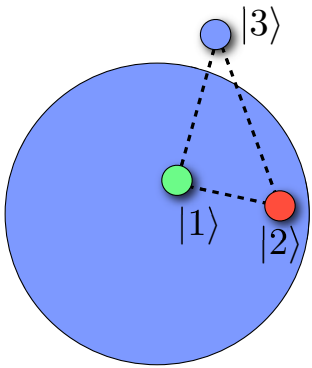}
\caption{A system with three distinct particles, $|1\rangle$, $|2\rangle$ and $|3\rangle$, with 
identical two-body interacting potentials in all subsystems (illustrated by the dashed lines). Particle
$|3\rangle$ has a Fermi sea (large (blue) filled sphere) that causes Pauli blocking of states in momentum 
space.}
\label{fig2}    
\end{figure*}

The first example we consider is the presence of a 
degenerate Fermi sea as a background. To simplify the 
problem we assume that we have three distinct particle 
and where one of the 
three feel the presence of a background of other particles 
of the same species. This implies that there will be 
Pauli blocking of the momentum states of this particle.
This is illustrated in Fig.~\ref{fig2}. In the 
Born-Oppenheimer limit, where the mass of the 
particle with the Fermi sea is much smaller than 
the masses of the two other particles, this problem 
was studied by Nishida \cite{nishida2009}, 
by MacNeill and Zhou \cite{macneill2011}, and 
more recently for several heavy particles 
by Endo and Ueda \cite{endo2013}. Here we 
will consider the case of equal masses, although we
note that the formalism discussed here can be applied
for any choice of masses and agrees with the Born-Oppenheimer
result in the appropriate limit.

Our approach will be top-down, i.e. we consider a formalism 
that can describe the Efimov states accurately and then 
introduce the Fermi sea background \cite{nygaard2011}. Given that a Fermi sea
is most easily described in momentum-space, we will work 
with the momentum-space formalism using the so-called
Skornyakov-Ter-Martirosian equations \cite{STM1956}, suitably
modified to avoid the Thomas collapse that is otherwise
present in the limit where the two-body interaction has zero range \cite{pricoupenko2011}.
Notice that we will consider the Fermi sea inert, i.e. it 
will not contain any fluctuations in the form of particle-hole
pairs. In the Born-Oppenheimer limit, it was argued \cite{macneill2011}
that fluctuations should not change the qualitative effect discussed here. 
Similar arguments can be made for general masses (this will 
be presented elsewhere). The key issue is to introduce
the proper Pauli blocking in the two-body propagator 
and then implement this in the three-body equations which 
are subsequently solved for negative (bound) energy states
using a pole and residue expansion \cite{braaten2006}.

\begin{figure*}
\centering
\includegraphics[scale=0.4]{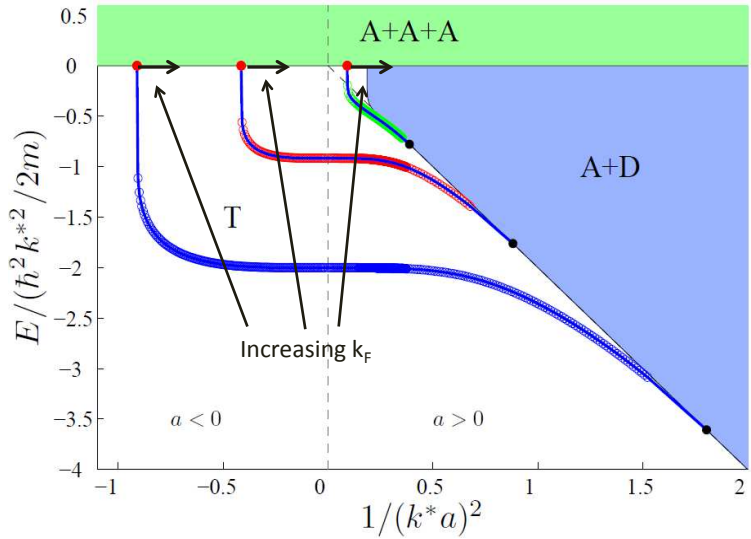}
\caption{The Efimov spectrum with a finite Fermi sea influencing one of the 
three particles that make up the three-body state. Both axis have been scaled by 
the fourth root for visibility. $k^*$ is a parameter that describes the 
short-range details of the interparticle interactions that are not captured
by a zero-range approximation. The three atom continuum (A+A+A) is at the top, 
the atom-dimer (A+D) continuum on the right side, while the trimer
states (T) live below both continua. As indicated, the presence of a finite
Fermi sea induces a spectral change or flow toward the atom-dimer continuum 
with increasing Fermi wave vector $k_F$.}
\label{fig3}     
\end{figure*}

In Fig.~\ref{fig3} we show the results of a three-body calculation with a 
finite Fermi sea background in one of the components. The spectrum demonstrates
the general modification of the binding energies in the presence of a 
background. For instance, the appearance points (filled (red) circles along the 
A+A+A continuum line) for Efimov three-body states are moved as the Fermi 
sea increases and eventually they are eaten by the atom-dimer (A+D) continuum. 
This can be understood intuitively by comparing the binding energy on 
resonance ($|a|\to\infty$), $E_B$, with the Fermi energy, $E_F$. When
$E_F<<E_B$ no spectral flow is seen, but once $E_F\sim E_B$ the states
start moving. In fact, we have found that this happens in a manner
that also displays Efimov scaling, but now in the Fermi energy or 
Fermi wave vector \cite{nygaard2011}; another state gets swallowed 
by the A+D continuum every time the Fermi wave vector scales by 
the Efimov factor, $e^\pi/s_0\sim 22.7$. This scaling factor can 
be changed by taking different masses. 

A prominent experimental feature of the Efimov effect is the 
loss rate as the three-body state merges with the 
three-atom continuum. As seen in Fig.~\ref{fig3}, our theory
predicts that these points are moving as a function of $E_F$. 
As discussed in Ref.~\cite{nygaard2011}, this implies that
the loss peak could be displaced due to the presence of the 
Fermi sea. An ideal system to study this effect is the 
three-component $^6$Li gas which has produced Efimov features in 
several experiments around the world \cite{huckans2009,jochim2010,fuji2011}.
Our calculations indicate that the peak should be affected
at densities of about $10^{12}$ cm$^{-3}$ and above. This is 
about one order of magnitude larger than the reported experiments
and could be within reach for this system. Of course, finite 
temperature effects or the unitarity bound (close to 
the resonance peaks are washed out) must be considered 
and further analysis of these effects is 
necessary.

\section{Efimov states in a Bose gas background}
Consider now the case where one of the particles in a three-body 
Efimov state has a Bose background. In the limit where $|a|\to\infty$, 
we again need to have some scale to compare to the three-body energy
in order to estimate whether we can expect an effect of the background. Some
recent discussions on the few- and many-body effects of the resonant 
Bose gas for equal mass bosons can be found in Refs.~\cite{borzov2012,zhou2013,jiang2013}.
In an ideal Bose gas we could construct an energy scale from the density, $n$, 
and the mass, $m$. A short moment of contemplation reveals that this 
scale is really $E_F$, and thus our estimates above would apply. However,
experimental alkali Bose gases typically have a short-range interaction 
between the identical bosons given by some scattering length, $a_B$. 
This implies that they will also have a superfluid length or coherence
length, $\xi=1/\sqrt{8\pi n a_B}$. This implies that for excitation 
with wave lengths longer than $\xi$ the collective properties of the Bose
condensate are dominant, or in terms of energy for modes with energies 
lower than about $\hbar^2/m\xi^2$. Since Efimov three-body states are 
low-energy states one may expect that a modification of the dispersion 
relation due to a Bose condensate background would have an effect.

In order to study this issue quantitatively, we have employed the 
Born-Oppenheimer approximation with a light particle Bose condensate
containing two heavy impurity atoms that interact with all the light 
bosons through a short-range interaction \cite{zinner2013}. This is 
similar to the Fermi gas study in Ref.~\cite{macneill2011}, but in the 
case of a Bose condensate background instead of Pauli blocking we get
a modified dispersion at low energy that has to be taken into account.
Again this is a top-down approach where we use a formalism that reproduces
the Efimov states when the condensate density goes to zero. In the 
Born-Oppenheimer limit, the effective potential for $|a|\to\infty$ 
(with $a$ the heavy-light scattering length) between the 
two heavy particles in 3D is 
\begin{equation}
V(R)=-\frac{\hbar^2\Omega^2}{mR^2},
\end{equation}
where $R$ is the distance between the two impurities, and $\Omega$ solves
the equation $e^{-\Omega}=\Omega$ so that $\Omega\sim 0.567$. This 
potential produces the Efimov effect with the 
number of three-body states given by 
\begin{equation}
N_T\approx \frac{1}{\pi}\textrm{Log}\left[\frac{a}{R_0}\right].
\end{equation}
Here $R_0$ is a short-range cuf-off which in the Born-Oppenheimer limit
is obtained from the short-range properties of the interaction of 
the two heavy particles (which is most likely given by the 
van der Waals length \cite{chin2011,ueda2012,wang2012,peder2012,schmidt2012}).
The presence of a Bose condensate modifies this formula 
by a replacement of $a$ by $\xi$, i.e.
\begin{equation}
N_T\approx \frac{1}{\pi}\textrm{Log}\left[\frac{\xi}{R_0}\right].
\end{equation}
This implies that the Efimov states are modified
at long distances (or low energy) by the deformation of the 
dispersion controlled by $\xi$. As discussed in Ref.~\cite{zinner2013}, 
mixtures of metastable $^4$He$^*$ and $^{87}$Rb \cite{borbely2012,knoop2012} could be 
a candidate system for observing this effect by a modification 
of the number of loss peaks as the condensate density is 
varied.

\section{Outlook}
The examples discussed above represent a first step in the exploration 
of the fate of Efimov states in a many-body environment. We assumed inert
backgrounds with no particle-hole pairs in the Fermi sea and no excitations
out of the condensate in the Bose condensate background case. Further
analysis has to be done to investigate the qualitative and quantitative
influence of fluctuations. This is similar in spirit to the 
Fermi polaron where a single spin down interacts with a Fermi sea of 
many spin up particles. There only single particle-hole pairs
are important due to higher order cancellations \cite{combescot2008}.
Scaling analysis of the Skornyakov-Ter-Martirosian equations 
with a Fermi sea suggest that at least the Efimov scaling with 
the Fermi energy $E_F$ should be robust when including fluctuations
but furhter studies on the quantitative effects are needed. 

The examples presented here are for 3D, and we would expect 
changes as we move to 2D setups. Here the Efimov effect
does not occur in a strict sense, but there are universal 
bound states in the zero-range interaction limit and they 
will be modified by background effects. Another interesting
direction would be Efimov three-body physics in a superfluid
Fermi gas where the spectrum is now gapped due to pairing. 
Finally, one may also ask what happens in the case of interactions
with longer ranges. For instance, for cold neutral heteronuclear molecules with
dipole-dipole interactions, few-body bound states are quite 
prolific and can have strong influence on the many-body ground 
state of the system 
\cite{pikovski2010,baranov2011,zinner2012,artem2013-dip}.
In 2D the two-body bound state is always
present for purely attractive potentials or dipole-dipole interactions
in layered geometries \cite{armstrong2012}. A goal would 
be to engineer a system with no bound two-body but 
bound three-body systems, a quantum trimer system.

\begin{acknowledgements}
I would like to thank Nicolai Gayle Nygaard for his pioneering work 
on the Efimov effect in the presence of a Fermi sea upon which much of this 
discussion is based. I would also like to thank Aksel S. Jensen, Dmitri
V. Fedorov, Tobias Frederico, and Marcelo Yamashita for discussions and 
for teaching me most of what I know about the Efimov effect. I also thank 
our students Peder K. S{\o}rensen, Artem G. Volosniev, and Filipe F. Bellotti
for their tireless work on various aspects of few-body physics and the Efimov
effect.
\end{acknowledgements}

\end{document}